\documentclass[article, nojss]{jss}

\usepackage[]{graphicx}
\usepackage[]{color}
\makeatletter
\def\maxwidth{ %
  \ifdim\Gin@nat@width>\linewidth
    \linewidth
  \else
    \Gin@nat@width
  \fi
}
\makeatother

\definecolor{fgcolor}{rgb}{0.345, 0.345, 0.345}
\newcommand{\hlnum}[1]{\textcolor[rgb]{0.686,0.059,0.569}{#1}}%
\newcommand{\hlstr}[1]{\textcolor[rgb]{0.192,0.494,0.8}{#1}}%
\newcommand{\hlcom}[1]{\textcolor[rgb]{0.678,0.584,0.686}{\textit{#1}}}%
\newcommand{\hlopt}[1]{\textcolor[rgb]{0,0,0}{#1}}%
\newcommand{\hlstd}[1]{\textcolor[rgb]{0.345,0.345,0.345}{#1}}%
\newcommand{\hlkwa}[1]{\textcolor[rgb]{0.161,0.373,0.58}{\textbf{#1}}}%
\newcommand{\hlkwb}[1]{\textcolor[rgb]{0.69,0.353,0.396}{#1}}%
\newcommand{\hlkwc}[1]{\textcolor[rgb]{0.333,0.667,0.333}{#1}}%
\newcommand{\hlkwd}[1]{\textcolor[rgb]{0.737,0.353,0.396}{\textbf{#1}}}%

\usepackage{framed}
\makeatletter
\newenvironment{kframe}{%
 \def\at@end@of@kframe{}%
 \ifinner\ifhmode%
  \def\at@end@of@kframe{\end{minipage}}%
  \begin{minipage}{\columnwidth}%
 \fi\fi%
 \def\FrameCommand##1{\hskip\@totalleftmargin \hskip-\fboxsep
 \colorbox{shadecolor}{##1}\hskip-\fboxsep
     \hskip-\linewidth \hskip-\@totalleftmargin \hskip\columnwidth}%
 \MakeFramed {\advance\hsize-\width
   \@totalleftmargin\z@ \linewidth\hsize
   \@setminipage}}%
 {\par\unskip\endMakeFramed%
 \at@end@of@kframe}
\makeatother

\definecolor{shadecolor}{rgb}{.97, .97, .97}
\definecolor{messagecolor}{rgb}{0, 0, 0}
\definecolor{warningcolor}{rgb}{1, 0, 1}
\definecolor{errorcolor}{rgb}{1, 0, 0}
\newenvironment{knitrout}{}{} 

\usepackage{alltt}

\usepackage[utf8]{inputenc}
\usepackage[english]{babel}
\usepackage[T1]{fontenc}
\usepackage{amsmath}
\usepackage{pstricks}

\usepackage{amsthm}
\usepackage{amssymb}
\usepackage{multirow}

\newcommand{\der}{\mbox{d}}

\newcommand{\KK}{\mathbb{K}}

\newcommand{\SSS}{\mathbb{S}}
\newcommand{\R}{\mathbb{R}}

\newcommand{\cF}{\mathcal{F}}
\newcommand{\cP}{\mathcal{P}}
\newcommand{\N}{\mathbb{N}}

\usepackage{multirow}

\usepackage{ulem} 

\def\id{\hbox{{\rm\bf 1}\kern-.4em\hbox{{\rm\bf 1}}}\! } 

\usepackage{dsfont}

\author{Justine Lequesne\\Centre Henri Becquerel  \And 
        Philippe Regnault\\Universit\'e de Reims Champagne-Ardenne}
\title{\pkg{vsgoftest}: An \proglang{R} Package for Goodness-of-Fit Testing Based on Kullback-Leibler Divergence}

\Plainauthor{Justine Lequesne, Philippe Regnault} 
\Plaintitle{Package vsgoftest for R: goodness-of-fit tests based on Kullback-Leibler divergence} 
\Shorttitle{\pkg{vsgoftest}: package for GOF testing based on KL-divergence} 

\Abstract{
The \proglang{R}-package \pkg{vsgoftest} performs goodness-of-fit (GOF) tests, based on Shannon entropy and Kullback-Leibler divergence, developed by \cite{vasicek} and \cite{song}, of various classical families of distributions. The theoretical framework of the so-called Vasicek-Song (VS) tests is summarized and followed by a detailed description of the different features of the package. 
The power and computational time performances of VS tests are studied through their comparison with other GOF tests. 
Application to real datasets illustrates the easy-to-use functionalities of the \pkg{vsgoftest} package.  
}
\Plainkeywords{R statistical computing environment, goodness-of-fit tests, Shannon entropy, Kullback-Leibler divergence, sample spacing based estimation}

\Address{
Justine Lequesne, Centre Henri Becquerel, Unit\'e de Recherche Clinique, Rue d'Amiens, CS 11516, 76038 Rouen 
cedex 1, France, \\
E-mail: \email{justine.lequesne@chb.unicancer.fr}\\

\vspace{0.2cm}
Philippe Regnault, Laboratoire de Math{\'e}matiques de Reims, FRE 2011, Universit\'e de Reims Champagne-Ardenne, Campus Moulin de la Housse, BP 1039, 51687 Reims cedex 2, France.\\
E-mail: \email{philippe.regnault@univ-reims.fr}
}

\IfFileExists{upquote.sty}{\usepackage{upquote}}{}
\begin{document}

\section{Introduction} \label{PKGSecIntro}

Goodness-of-fit (GOF) tests constitute a classical tool in deciding of the compatibility of data with a theoretical (probability) distribution. The present work proposes a package for the \proglang{R} statistical computing environment~\cite{Rcran} performing GOF tests based on Shannon entropy and Kullback-Leibler divergence, together with a methodological guide and applications. 

Precisely, we consider fitting numeric (real valued) data either to a unique distribution, the so-called simple null hypothesis test
\begin{equation}\label{arthyps}
H_{0}: P ={P}_{0}(\theta) \quad\mbox{against}\quad
H_{1}: P \ne {P}_{0}(\theta),
\end{equation}
or to a parametric family, the so-called composite null hypothesis test
\begin{equation}\label{arthyp}
H_{0}: P \in \cP_{0}(\Theta) \quad\mbox{against}\quad
H_{1}: P \notin \cP_{0}(\Theta).
\end{equation}
The set $\cP_{0}(\Theta)=\{ P\in\mathcal{D} : P=P_{0}(\theta), \theta \in \Theta  \}$, with $\Theta \subset \overline{\R}^{d}$, is a parametric subfamily of the set $\mathcal{D}$ of all probability distributions absolutely continuous with respect to Lebesgue measure on $\R$, \textit{i.e.}, probability distributions with a density function.
Decision is to be taken from the observation $x_1^n=(x_{1}, \ldots, x_{n})$  of
a sample  $X_1^n=(X_1,\dots,X_n)$ of size $n$ of independent and identically distributed random variables drawn from $P \in \mathcal{D}$.

Classically, a GOF test procedure is derived by computing some distance-like functional between the observations and the null distribution, or family of distributions, the null hypothesis being rejected when the distance is larger than a critical value.
Kolmogorov-Smirnov, Cram\'er-von Mises and Anderson-Darling tests constitute some of the most commonly used GOF tests. 
Their test statistics measure discrepancy  between the empirical cumulative distribution function of the sample and the cumulative distribution function of the null distribution; these tests are refered to as EDF tests in the following; see~\cite{stephens1974edf}.
The tests in this paper are based on the Kullback-Leibler (KL) divergence of the density of the sample with respect to the null density. 

GOF tests based on KL divergence have been introduced by \cite{vasicek} for testing normality. 
Vasicek normality test relies on the maximum entropy property satisfied by the normal distribution: amoung all distributions with density and finite variance, Shannon entropy is maximized by the normal distribution. 
Vasicek test statistic is a monotone function of the entropy difference between the null normal distribution and the observed one.
This has been subsequently extended to GOF tests of uncategorical data for numerous families of distributions satisfying a maximum entropy property -- say maximum entropy (ME) distributions; see Section~\ref{MEGOFPSecFramework} for references. GOF tests based on entropy differences are known to have higher power than classical GOF tests in numerous cases; see Section~\ref{seccomputation}.
\cite{song} considers GOF tests based on KL divergence, for a large class of distributions including all classical distribution families. The test statistic is an estimate of the KL divergence between the sample and the null distributions. It is asymptotically normally distributed. When applied to {ME} distributions, it is equal to the difference between the entropies of the null distribution and the sample one, yielding the same decision rule as Vasicek test. 
This paper presents the implementation of Vasicek and Song tests (VS tests) for various families of distributions: uniform, normal, log-normal, exponential, gamma, Weibull, Pareto, Fisher, Laplace and Beta distributions. For further details on the theoretical aspects of VS tests, see \cite{GirLeq2016_CSTM_EntropyBasedGOFTests} in which a unifying framework for tests based on entropy difference and KL divergence is provided.

Numerous \proglang{R} packages perform GOF tests for various families of distributions. The functions \code{chisq.test}, \code{ks.test} and \code{shapiro.test} of the \pkg{stats} package perform respectively the chi-squared test of adequacy to a discrete distribution, the Kolmogorov-Smirnov GOF test for any theoretical continuous distribution and the Shapiro-Wilk normality test. 
The packages \pkg{goftest} developed in~\cite{goftest_package} and \pkg{goft} in~\cite{goft_package} perform respectively Cram\'er-von Mises and Anderson-Darling GOF tests, and tests based on the ratios of variance and other moment estimators. \pkg{KScorrect} in~\cite{KScorrect_package} performs the Lilliefors-corrected Kolmogorov-Smirnov GOF test. Numerous GOF tests of the exponential or two-parameter Weibull distributions are  available in \pkg{EWGoF}~\cite{EWGoF_package} while \pkg{nortest} and \pkg{normtests} are dedicated to testing normality. 
The \pkg{dbEmpLikeGOF} package developed in~\cite{dbEmpLikeGOF_package} proposes GOF normality and uniformity tests based on empirical likelihood ratio. These tests are closely related with VS tests; similarities and differences between them are highlighted in the following sections.  

The test procedure implemented in \pkg{vsgoftest} uses either the asymptotic distribution of the test statistic or Monte-Carlo simulation, depending on sample size or user's choice. 
Optional arguments are included for handling particular situations such as samples with numerous ties. They also contribute to make the procedure flexible and fully parameterizable. Besides these practical aspects, the paper presents a comprehensive review of the literature dealing with power properties of VS tests. Monte Carlo simulations are conducted to illustrate their performance when applied to discriminate between close distributions.

The paper is organized as follows. The theoretical framework of GOF tests based on Shannon entropy difference and KL divergence is briefly presented in Section~\ref{MEGOFPSecFramework}. The functionalities of \pkg{vsgoftest} are presented in Section~\ref{MEGOFPSecPackage}. The tests performed by \pkg{vsgoftest} are compared to other GOF tests in Section~\ref{MEGOFTSecComparison}. More precisely, power comparisons to classical GOF tests are presented in Section~\ref{seccomputation}; Section~\ref{PKGSecCompPkg} focuses on the comparison of \pkg{vsgoftest} and \pkg{dbEmpLikeGOF} test procedures, which rely on very close theoretical frameworks but significantly differ in some of their features. Finally, applications to real data in Section~\ref{MEGOFTSecApplication} illustrate the  usage of the proposed functionalities.

\section{Entropy difference and KL divergence based GOF tests} 
\label{MEGOFPSecFramework}

The Shannon entropy of a distribution $P$ with density function $p$ on $\R$ has been defined in \citet*{shannon} as
\begin{equation}
\SSS(P):=-\int_\R p(x) \log p(x) \der x.
\label{shannon}
\end{equation}
Entropy measures the uncertainty or variability of a distribution. The maximum entropy principle under moment constraints, or  ME method, favours  distributions with highest entropy for their highest degree of uncertainty; see \citet*{shannon} and \citet*{jaynes}. 
Among all distributions supported by a given finite length interval $I$ in $\R$, entropy is maximum and equals~$\log |I|$ for the uniform distribution, where~$|I|$ denotes the length of~$I$. Hence, the entropy difference $\log |I| - \SSS(P)$ can be thought as a distance-like measure between~$P$ and the uniform distribution.

Similarly, among all continuous distributions  supported within $\R$ with mean $\mu$ and variance~$\sigma^2$, Shannon entropy is maximum for the normal distribution $\mathcal{N}(\mu, \sigma^2)$ and equals 
\begin{equation} \label{MEGOFPEqnEntNormal}
 \SSS(\mathcal{N}(\mu, \sigma^2)) = \ln(\sigma \sqrt{2\pi e}).
\end{equation}
The entropy difference $\SSS(\mathcal{N}(\mu, \sigma^2)) - \SSS(P)$ is nonnegative and thus defines a distance-like  measure between any distribution with mean $\mu$ and variance $\sigma^2$ and the $\mathcal{N}(\mu, \sigma^2)$ distribution. Based on this property, \citet*{vasicek} derives a normality test, with a test statistic expressed in terms of entropy differences, defined as follows 
$$K_{mn}:=\frac{n}{2mS}\left\lbrace \prod_{i=1}^{n} (X_{(i+m)}-X_{(i-m)}) \right\rbrace^{1/n}
=\sqrt{2\pi e}\exp \left(V_{mn}-\SSS \left( \mathcal{N} \left(\overline{X},S^2 \right) \right) \right),$$
where 
$$\overline{X}:=\frac{1}{n}\sum\limits_{i=1}^{n} X_{i} \quad \mbox{and} \quad S^2:= \frac{1}{n}\sum\limits_{i=1}^{n}(X_i-\overline{X})^2 $$
are the empirical estimators of respectively the mean and the variance of the sample $X_1^n:=(X_1, \dots, X_n)$, $X_{(1)} \leq \dots \leq X_{(n)}$ denotes the order statistics associated to $X_1^n$ and
\begin{equation} \label{MEGOFPEqnVasiEst}
 V_{mn} :=\frac{1}{n} \sum_{i=1}^{n} \log \left( \frac{n}{2m} \left[ X_{(i+m)}-X_{(i-m)} \right] \right)
\end{equation}
is the non-parametric Vasicek estimator of $\SSS(P)$ based on spacings, with $X_{(i)}=X_{(1)}$ if $i<1$ and  $X_{(i)}=X_{(n)}$ if $i>n$; the window size $m\in\N^*$ is smaller than $n/2$. 

The test statistic has been extended to various families of ME distributions under moment constraints; see \citet*{dudewicz},
\citet*{ebrahimi}, \citet*{choikim}, \citet*{mergel}, \citet*{mudholkar}, among many others.  
A unifying framework for any exponential family of distributions is proposed in~\cite{GirLeq2016_CSTM_EntropyBasedGOFTests}, with asymptotic properties, consistency and application to biology; see also~\cite{lequesne_maxent}, \cite{lequesne_these} and \cite{lequesne_maxent2} for power efficiencies, GOF tests of Pareto distributions and extension to generalized entropies.

The Kullback-Leibler (KL) divergence of a distribution $P$ with respect to another one $Q$, is defined as 
\begin{equation}
 \KK(P| Q):=\int_\R p(x) \log \frac{p(x)}{q(x)} \der x,
 \label{kullback}
 \end{equation}
if $P$ is absolutely continuous with respect to $Q$, with respective densities $p$ and $q$, and as
$+\infty$ if not; see \citet*{kullback}. The KL divergence is linked to Shannon entropy through the relation 
\begin{equation} \label{PKGEqnRelKLSE}
\KK(P | Q)=- \SSS(P)-\int_{\R} p(x) \log q(x) \der x.
\end{equation}
The KL divergence is not a mathematical distance because of lack of both symmetry and triangular inequality, but it satisfies $\KK(P | Q) \geq 0,$ with $\KK(P | Q)=0$ if and only if $P=Q$, and thus constitutes a natural measure of discrepancy for GOF tests. 
\citet*{song} proposes GOF tests based on KL divergence for either simple (\ref{arthyps}) or composite (\ref{arthyp}) null hypothesis. Precisely, thanks to~(\ref{PKGEqnRelKLSE}), the test statistic $I_{mn}$ is the estimator of $\KK(P | P_{0}(\theta))$, defined by
\begin{equation} \label{artqimn}
 I_{mn} := -V_{mn}-\frac{1}{n} \sum_{i=1}^{n} \log p_{0}(X_{i},{\widehat{\theta}}_{n}),
\end{equation}
where $V_{mn}$ is the Vasicek estimator~(\ref{MEGOFPEqnVasiEst}) of $\SSS(P)$, and $\widehat{\theta}_{n}$ is either the maximum likelihood estimator (MLE) of $\theta$ satisfying
$$\frac{1}{n}\sum_{i=1}^n \log p_{0}(X_{i},{\widehat{\theta}}_{n}) = \max_{\theta \in \Theta} \frac{1}{n} \sum_{i=1}^{n} \log p_{0}(X_{i},\theta),$$
or $\theta$ itself in case of a simple null hypothesis~(\ref{arthyps}).

The KL divergence $\KK(P|Q)$ for a maximum entropy distribution $Q$ under moment constraints reduces to the entropy difference $\SSS(Q) - \SSS(P)$ for all $P$ satisfying the same moment constraints; see~\cite{csiszar}. 
This Pythagorean equality allowed \cite{GirLeq2016_CSTM_EntropyBasedGOFTests} to establish that entropy difference GOF test for ME distributions coincide with Song test -- we will refer to these tests as Vasicek-Song tests and keep on denoting them by VS tests. Especially, Vasicek and Song normality test statistics are linked through the equality
$$K_{mn} = \sqrt{2\pi e} \exp(-I_{mn}),$$
yielding identical decision rules.

Based on the asymptotic properties of $V_{mn}$ proven by \citet*{dudewicz} for testing uniformity, \citet*{song} establishes the asymptotic behavior of $I_{mn}$, independently of the null hypothesis: $I_{mn}$ is consistent and asymptotically normally distributed provided the null distribution belongs to the class 
\begin{equation}
\cF=\left\{ P\in\mathcal{D}: \sup_{x:\; 0<F(x)<1} \frac{\vert p'(x) \vert}{p^{2}(x)} F(x)[1-F(x)] < \gamma, \ \gamma >0 \right\},
 \label{artfamilleasymptotique}
 \end{equation}
where $F$  is the cumulative distribution function of $P$ and $p$ its density with derivative $p'$ (almost everywhere).
The class $\cF$ contains the most classical distributions such as uniform ($\gamma=0$), normal, exponential and gamma ($\gamma=1$), Fisher ($\gamma=(2+\nu_2)/\nu_2$ where $\nu_2$ is the second degree of freedom), Pareto ($\gamma=(\mu+1)/\mu$, where $\mu$ is the shape parameter), etc. 
For $\cP_0(\Theta)\subset\cF$, if 
\begin{equation} \label{PKGEqnConditionsWindow}
m/\log n \xrightarrow[n \rightarrow \infty]{} 0 \quad \textrm{and} \quad m(\log n)^{2/3}/n^{1/3} \xrightarrow[n \rightarrow \infty]{} 0,
\end{equation}
then
\begin{equation} \label{asymptoticn}
\sqrt{6mn}[I_{mn}-\log(2m)+\psi(2m)] \xrightarrow{\cal D} \mathcal{N}(0,1),
\end{equation}
where $\psi(m)$ is the digamma function.
The asymptotic bias  $\log (2m)-\psi(2m)$ of $I_{mn}$ is that  of $-V_{mn}$. \citet*{song} suggests a bias correction in the asymptotic distribution ~(\ref{asymptoticn}) for moderate sample sizes:
\begin{equation} \label{asymptoticn2}
\sqrt{6mn}\left[ I_{mn}-b_{mn} \right] \xrightarrow{\cal D} \mathcal{N}(0,1),
\end{equation}
where 
$$b_{mn} := \log(2m) - \log(n) -\psi(2m) + \psi(n+1) +\frac{2m}{n} R_{2m-1} - \frac{2}{n} \sum_{i=1}^m R_{i+m-2},$$
with $R_{m} := \sum_{j=1}^m 1/j$.
From (\ref{asymptoticn2}), an asymptotic p-value for the related VS test is given by
\begin{equation} \label{asymppvalue}
p=1-\Phi^{-1}\left( \sqrt{6mn}\left[I_{mn}(x_1^n)-b_{mn}\right]\right),
\end{equation}
where $I_{mn} (x_1^n)$ denotes the value of the statistic $I_{mn}$ for the observations $x_1^n = (x_1, \dots, x_n)$, and $\Phi$ denotes the cumulative distribution function of the normal distribution. According to~\cite{song}, the asymptotic p-value (\ref{asymppvalue}) provides accurate results for sample sizes $n$ larger than 80. For small sample sizes, Monte Carlo simulations should be preferred. 
A large number~$N$ of replications of $X_1^n$  drawn from $P_{0}(\widehat{\theta}_{n})$ (or $P_{0}(\theta)$ in case of simple null hypothesis) are generated.
The test statistic $I_{mn}^{i}$ is computed for each replication $i, 1\leq i\leq N.$ 
The p-value is then given by the empirical mean $(\sum_{i=1}^{N}\mathds{1}_{\{I_{mn}^{i}>I_{mn}(x_1^n)\}})/N.$

For choosing $m$, \citet*{song} proposes to minimize $I_{mn}$  -- that is to maximize $V_{mn}$, with respect to~$m$, yielding the most conservative test. The KL divergence $\KK(P |P_0(\theta))$ being nonnegative, values of $m$ for which $I_{mn}$ is negative are excluded, leading to choose $m$ subject to the constraint
\begin{equation} \label{PKGEqnConstraint}
V_{mn} \leq - \frac{1}{n} \sum_{i=1}^n \log  p_{0} \left(.; \widehat{\theta}_n \right).
\end{equation}
Finally, the window size proposed by \citet*{song} -- say the optimal window size, is
\begin{equation}\label{arteqm}
\widehat{m} := \min
   \left\{
    m^* \in \arg\!\!\max_{m\in\N^*} 
    \left\{ 
      V_{mn} : \; V_{mn}\leq -\frac{1}{n} \sum_{i=1}^{n} \log p_{0} \left( X_{i}, \widehat{\theta}_{n} \right)
    \right\}
    : 1 \leq m^{*} < \lfloor n^{1/3-\delta} \rfloor
   \right\},
\end{equation}
for some $\delta< 1 /3$ and the VS test statistic is then
\begin{equation} \label{PKGEqnTestStat}
 I_{\widehat{m}n} = - V_{\widehat{m}n} - \frac{1}{n} \sum_{i=1}^n \log p_0 \left( X_i;\widehat{\theta}_n \right).
\end{equation}
The upper bound $n^{1/3 - \delta}$ for the window size $m$ is chosen so that conditions~(\ref{PKGEqnConditionsWindow}) are fulfilled and hence that asymptotic normality~(\ref{asymptoticn}) holds. No optimal choice of $\delta$ exists; it depends on the family of distributions of the null hypothesis; see Section~\ref{MEGOFPSecPackage} for details.

The package \pkg{vsgoftest} presented below performs VS GOF tests for several parametric families of ME distributions: uniform, normal, log-normal, exponential, Pareto, Laplace, Weibull, Fisher, gamma and beta distributions. These families of distribution, all included in the class~$\cF$ given by~(\ref{artfamilleasymptotique}), have been chosen so that the package covers a large variety of applications. 
Note that the package \pkg{dbEmpLikeGOF} performs uniformity and normality VS tests, with an alternative choice for the window size. Precisely, the test statistic is $nI_{mn} + 1/2$, and the window $m$ is chosen, between $1$ and $n^{1/2}$, minimizing $nI_{mn}$. The constraint~(\ref{PKGEqnConstraint}) is not considered. 
The asymptotic distribution of $nI_{mn}$ is not used, p-values being computed from a pre-calculated table for small sample sizes or via Monte-Carlo simulation; see~\cite{dbEmpLikeGOF_package} and~\cite{VexlerGurevich2010_CSDA}.
This alternative methodological approach leads to different decisions that may be less reliable, particularly when applied to heavy tailed samples. Other differences in the coding structure make \pkg{vsgoftest} faster than \pkg{dbEmpLikeGOF}, especially when Monte-Carlo simulation is performed. These points will be detailled in Section~\ref{PKGSecCompPkg}.

\section{The package vsgoftest} 
\label{MEGOFPSecPackage}

The \pkg{vsgoftest} package provides functions for estimating Shannon entropy of absolutely continuous distributions and testing the goodness-of-fit of some theoretical family of distributions to a vector of real numbers. It also provides functions for computing the density, cumulative density and quantile functions of Pareto and Laplace distributions, as well as for generating samples from these distributions. 

The \pkg{vsgoftest} package is available on CRAN mirrors and can be installed by executing the command

\begin{knitrout}
\definecolor{shadecolor}{rgb}{0.969, 0.969, 0.969}\color{fgcolor}\begin{kframe}
\begin{alltt}
\hlkwd{install.packages}\hlstd{(}\hlstr{'vsgoftest'}\hlstd{)}
\end{alltt}
\end{kframe}
\end{knitrout}

Alternatively, the latest (under development) version of the \pkg{vsgoftest} package is also available and can be installed in \proglang{R} from the github repository of the project as follows:

\begin{knitrout}
\definecolor{shadecolor}{rgb}{0.969, 0.969, 0.969}\color{fgcolor}\begin{kframe}
\begin{alltt}
\hlcom{#Package devtools must be installed}
\hlstd{devtools}\hlopt{::}\hlkwd{install_github}\hlstd{(}\hlkwc{repo} \hlstd{=} \hlstr{'pregnault/vsgoftest'}\hlstd{)}
\end{alltt}
\end{kframe}
\end{knitrout}

The package is structured around two functions, \code{entropy.estimate} and \code{vs.test}; the first one computes the spacing based estimator~(\ref{MEGOFPEqnVasiEst}) from a numeric sample, the second one performs Vasicek-Song GOF test for usual parametric families of distributions based on the test statistic~(\ref{PKGEqnTestStat}). 
A comprehensive presentation of their usage is proposed in Sections~\ref{MEGOFPSecEntEst} and~\ref{MEGOFPSecVSTest}, with numerous examples. Section~\ref{MEGOFPSecStrPack} provides further technical information about the structure of the package.

\subsection{Function entropy.estimate for estimating Shannon entropy} \label{MEGOFPSecEntEst}

The  function \code{entropy.estimate} computes the spacing based estimate~(\ref{MEGOFPEqnVasiEst}) of Shannon entropy~(\ref{shannon}) from a numeric sample. Two arguments have to be provided:
\begin{itemize}
 \item \code{x}: the numeric sample;
 \item \code{window}: an integer between 1 and half of the sample size, specifying the window size of the spacing-based estimator~(\ref{MEGOFPEqnVasiEst}).
\end{itemize}
It returns the estimate of Shannon entropy of the sample. Here is an example for a sample drawn from a normal distribution with parameters $\mu = 0$ and $\sigma^2 = 1$.

\begin{knitrout}
\definecolor{shadecolor}{rgb}{0.969, 0.969, 0.969}\color{fgcolor}\begin{kframe}
\begin{alltt}
\hlkwd{library}\hlstd{(}\hlstr{'vsgoftest'}\hlstd{)}
\end{alltt}

{\ttfamily\noindent\itshape\color{messagecolor}{Loading required package: fitdistrplus}}

{\ttfamily\noindent\itshape\color{messagecolor}{Loading required package: MASS}}

{\ttfamily\noindent\itshape\color{messagecolor}{Loading required package: survival}}\begin{alltt}
\hlkwd{set.seed}\hlstd{(}\hlnum{2}\hlstd{)}     \hlcom{#set seed of PRNG}
\hlstd{samp} \hlkwb{<-} \hlkwd{rnorm}\hlstd{(}\hlkwc{n} \hlstd{=} \hlnum{100}\hlstd{,} \hlkwc{mean} \hlstd{=} \hlnum{0}\hlstd{,} \hlkwc{sd} \hlstd{=} \hlnum{1}\hlstd{)} \hlcom{#sampling from normal distribution}
\hlkwd{entropy.estimate}\hlstd{(}\hlkwc{x} \hlstd{= samp,} \hlkwc{window} \hlstd{=} \hlnum{8}\hlstd{)} \hlcom{#estimating entropy with window = 8}
\end{alltt}
\begin{verbatim}
[1] 1.394728
\end{verbatim}
\begin{alltt}
\hlkwd{log}\hlstd{(}\hlnum{2}\hlopt{*}\hlstd{pi}\hlopt{*}\hlkwd{exp}\hlstd{(}\hlnum{1}\hlstd{))}\hlopt{/}\hlnum{2} \hlcom{#the exact value of entropy}
\end{alltt}
\begin{verbatim}
[1] 1.418939
\end{verbatim}
\end{kframe}
\end{knitrout}

The estimate returned by \code{entropy.estimate} obviously depends on the window selected by the user, as illustrated by the following chunck.

\begin{knitrout}
\definecolor{shadecolor}{rgb}{0.969, 0.969, 0.969}\color{fgcolor}\begin{kframe}
\begin{alltt}
\hlkwd{sapply}\hlstd{(}\hlnum{1}\hlopt{:}\hlnum{10}\hlstd{,} \hlkwa{function}\hlstd{(}\hlkwc{w}\hlstd{)} \hlkwd{entropy.estimate}\hlstd{(}\hlkwc{x} \hlstd{= samp,} \hlkwc{window} \hlstd{=w))}
\end{alltt}
\begin{verbatim}
 [1] 1.205018 1.346352 1.378732 1.387337 1.391691 1.393512 1.394428
 [8] 1.394728 1.394486 1.392669
\end{verbatim}
\end{kframe}
\end{knitrout}

One may select the window size that maximizes the entropy estimate, as follows.

\begin{knitrout}
\definecolor{shadecolor}{rgb}{0.969, 0.969, 0.969}\color{fgcolor}\begin{kframe}
\begin{alltt}
\hlstd{n} \hlkwb{<-} \hlnum{100} \hlcom{#sample size}
\hlstd{V} \hlkwb{<-} \hlkwd{sapply}\hlstd{(}\hlnum{1}\hlopt{:}\hlstd{(n}\hlopt{/}\hlnum{2} \hlopt{-} \hlnum{1}\hlstd{),} \hlkwa{function}\hlstd{(}\hlkwc{w}\hlstd{)} \hlkwd{entropy.estimate}\hlstd{(}\hlkwc{x} \hlstd{= samp,} \hlkwc{window} \hlstd{=w))}
\hlkwd{which.max}\hlstd{(V)} \hlcom{#Choose window that maximizes entropy}
\end{alltt}
\begin{verbatim}
[1] 8
\end{verbatim}
\end{kframe}
\end{knitrout}

Let us consider a sample drawn from a Pareto distribution with density
$$p(x;c,\mu) = \frac{\mu c^\mu}{x^{\mu+1}}, \quad  x \geq c,$$
where $c >0 $ and $\mu >0 $, which can be obtained by making use of the function \code{rpareto} as illustrated below. Its Shannon entropy is 
$$\SSS(p( . ;c, \mu)) = -\ln \mu + \ln c + \frac{1}{\mu} +1.$$

\begin{knitrout}
\definecolor{shadecolor}{rgb}{0.969, 0.969, 0.969}\color{fgcolor}\begin{kframe}
\begin{alltt}
\hlkwd{set.seed}\hlstd{(}\hlnum{5}\hlstd{)}
\hlstd{n} \hlkwb{<-} \hlnum{100} \hlcom{#Sample size}
\hlstd{samp} \hlkwb{<-} \hlkwd{rpareto}\hlstd{(n,} \hlkwc{c} \hlstd{=} \hlnum{1}\hlstd{,} \hlkwc{mu} \hlstd{=} \hlnum{2}\hlstd{)} \hlcom{#sampling from Pareto distribution}
\hlkwd{entropy.estimate}\hlstd{(}\hlkwc{x} \hlstd{= samp,} \hlkwc{window} \hlstd{=} \hlnum{3}\hlstd{)}
\end{alltt}
\begin{verbatim}
[1] 0.8480204
\end{verbatim}
\begin{alltt}
\hlopt{-}\hlkwd{log}\hlstd{(}\hlnum{2}\hlstd{)} \hlopt{+} \hlnum{3}\hlopt{/}\hlnum{2} \hlcom{#Exact value of entropy}
\end{alltt}
\begin{verbatim}
[1] 0.8068528
\end{verbatim}
\end{kframe}
\end{knitrout}

\subsection{Function vs.test for testing GOF to a specified model} \label{MEGOFPSecVSTest}

The function \code{vs.test} performs the VS test, as described in Section~\ref{MEGOFPSecFramework}; setting two non-optional arguments is required:
\begin{itemize}
 \item \code{x}: the numeric sample;
 \item \code{densfun}: a character string specifying the theoretical family of distributions of the null hypothesis.  Available families of distributions are: uniform, normal, log-normal, exponential, gamma, Weibull, Pareto, Fisher and Laplace distributions. They are referred to by the symbolic name in \proglang{R} of their density function. For example, set \code{densfun = 'dnorm'} to test GOF of the family of normal distributions; see Table~\ref{tab:lois} for details.
\end{itemize}

It returns an object of class \code{htest}, \textit{i.e.}, a list whose main components are:
\begin{itemize}
 \item \code{statistic}: the value of VS test statistic~(\ref{PKGEqnTestStat}) for the sample, with optimal window size defined by~(\ref{arteqm});
 \item \code{parameter}: the optimal window size;
 \item \code{estimate}: the maximum likelihood estimate of the parameters of the null distribution (for the test~(\ref{arthyp}) with composite null hypothesis);
 \item \code{p.value}: the p-value associated to the sample.
\end{itemize}

By default, \code{vs.test} performs the composite VS test  of the family of distributions \code{densfun} for the sample \code{x}. The p-value is estimated by means of Monte-Carlo simulation if the sample size is smaller than 80, or through the asymptotic distribution~(\ref{asymptoticn}) of the VS test statistic otherwise.

\renewcommand{\arraystretch}{1.5}
\begin{table}[ht!]
 \centering
 \scalebox{0.9}{
 \small
 \begin{tabular}{|c|c|c|c|c|}
 \hline
 \textbf{Distribution} & \textbf{Call} (\code{densfun}) & \textbf{Parameters} & \textbf{Density} &   \textbf{Default $\delta$}\\
 \hline \hline
  Uniform & \code{"dunif"} & {$a<b$} & $\displaystyle \frac{1}{b-a} \id_{[a,b]}(x)$ & $1/12$\\ \hline
  Normal & \code{"dnorm"} &{ $\mu\in\R, \sigma\in\R_+$ } & $\displaystyle \frac{1}{\sqrt{2\pi} \sigma} \exp\left( - \frac{1}{2\sigma^2} (x-\mu)^2 \right)$ & {$1/12$}\\ \hline
  Log-normal& \code{"dlnorm"} & $\mu\in\R, \sigma\in\R_+$ & $\frac{1}{x \sqrt{2\pi}\sigma} \exp\left(- \frac{(\log x - \mu)^2}{2\sigma^2} \right) \id_{\mathbb{R}_+}(x)$ & $1/12$ \\  \hline
  Exponential& \code{"dexp"}& $\lambda\in\R$ & $\lambda \exp (- \lambda x)\id_{\mathbb{R}_+}(x)$ & $1/12$\\ \hline
  Pareto& \code{"dpareto"} & { $c\in\R,\mu\in\R_+$} & $\displaystyle \mu c^{\mu}\frac{1}{x^{\mu+1}}\id_{[c, \infty[}(x)$ & $1/12$\\ \hline
  Laplace& \code{"dlaplace"} & { $\mu\in\R,\sigma\in\R_+$ }& { $\displaystyle \frac{1}{2\sigma} \exp \left( - \frac{\rvert x-\mu\rvert}{\sigma}\right)$ }& $1/12$\\  \hline
  Weibull& \code{"dweibull"} & $a,b\in\R_+$&$\displaystyle \frac{a}{b^a} x^{a-1} \exp\left[ - \left(\frac{x}{b}\right)^a \right] \id_{\mathbb{R}_+}(x)$ & $2/15$ \\ \hline
  Fisher& \code{"df"} &$\nu_1,\nu_2\in\R_+$&$\displaystyle\frac{\left(\frac{d_1 x}{d_1 x+d_2}\right)^{d_{1}/2} \left(1-\frac{d_1 x}{d_1 x+d_2}\right)^{d_{2}/2}}{xB(d_{1}/2,d_{2}/2)} \id_{\mathbb{R}_+}(x)$& $2/15$\\ \hline
  Gamma& \code{"dgamma"}  &$\alpha,\beta \in\R_+$ &$\displaystyle\frac{\beta^{\alpha}}{\Gamma(\alpha)}x^{\alpha-1}\exp(-\beta x) \id_{\mathbb{R}_+}(x)$ & $2/15$\\ \hline
  Beta& \code{"dbeta"} & $\alpha, \beta \in\R$ & $\displaystyle \frac{x^{\alpha-1}(1-x)^{\beta-1}}{B(\alpha,\beta)} \id_{\mathbb{R}_+}(x)$ & $2/15$\\ \hline
 \end{tabular}
 }
  \caption{Families of distributions supported as the null by the \pkg{vsgoftest} package. The column denoted ``Default $\delta$" corresponds to the default setting for the parameter $\delta$; see~(\ref{arteqm}). Note that $B(a,b)= \int_0^1 x^{a-1}(1-x)^{b-1} dx$ is the Beta function.}
  \label{tab:lois}
\end{table}

In the following example, a normally distributed sample is simulated. 
VS test rejects the null hypothesis that this sample is drawn from a Laplace distribution, but does not reject the normality hypothesis (for a significant level set to $0.05$).

\begin{knitrout}
\definecolor{shadecolor}{rgb}{0.969, 0.969, 0.969}\color{fgcolor}\begin{kframe}
\begin{alltt}
\hlkwd{set.seed}\hlstd{(}\hlnum{5}\hlstd{)}
\hlstd{samp} \hlkwb{<-} \hlkwd{rnorm}\hlstd{(}\hlnum{50}\hlstd{,}\hlnum{2}\hlstd{,}\hlnum{3}\hlstd{)}
\hlkwd{vs.test}\hlstd{(}\hlkwc{x} \hlstd{= samp,} \hlkwc{densfun} \hlstd{=} \hlstr{'dlaplace'}\hlstd{)}
\end{alltt}
\begin{verbatim}

	Vasicek-Song GOF test for the Laplace distribution

data:  samp
Test statistic = 0.32437, Optimal window = 2, p-value = 0.0248
sample estimates:
   Shape    Scale 
2.194803 2.687321 
\end{verbatim}
\end{kframe}
\end{knitrout}

\begin{knitrout}
\definecolor{shadecolor}{rgb}{0.969, 0.969, 0.969}\color{fgcolor}\begin{kframe}
\begin{alltt}
\hlkwd{set.seed}\hlstd{(}\hlnum{4}\hlstd{)}
\hlkwd{vs.test}\hlstd{(}\hlkwc{x} \hlstd{= samp,} \hlkwc{densfun} \hlstd{=} \hlstr{'dnorm'}\hlstd{)}
\end{alltt}
\begin{verbatim}

	Vasicek-Song GOF test for the normal distribution

data:  samp
Test statistic = 0.21655, Optimal window = 2, p-value = 0.3704
sample estimates:
    Mean St. dev. 
2.194803 3.173824 
\end{verbatim}
\end{kframe}
\end{knitrout}

For performing a simple null hypothesis GOF test, the additional argument \code{param} has to be set to a numeric vector, consistent with the parameter requirements for the null distribution.
In such case, the MLE of the parameter(s) of the null distribution has not to be computed and hence the component \code{estimate} in results is not available.

\begin{knitrout}
\definecolor{shadecolor}{rgb}{0.969, 0.969, 0.969}\color{fgcolor}\begin{kframe}
\begin{alltt}
\hlkwd{set.seed}\hlstd{(}\hlnum{26}\hlstd{)}
\hlkwd{vs.test}\hlstd{(}\hlkwc{x} \hlstd{= samp,} \hlkwc{densfun} \hlstd{=} \hlstr{'dnorm'}\hlstd{,} \hlkwc{param} \hlstd{=} \hlkwd{c}\hlstd{(}\hlnum{2}\hlstd{,}\hlnum{3}\hlstd{))}
\end{alltt}
\begin{verbatim}

	Vasicek-Song GOF test for the normal distribution with Mean=2,
	St. dev.=3

data:  samp
Test statistic = 0.22196, Optimal window = 2, p-value = 0.331
\end{verbatim}
\end{kframe}
\end{knitrout}

If \code{param} is not consistent with the specified distribution -- e.g., standard deviation for testing a normal distribution is missing or negative, the execution is stopped and an error message is returned.

{
  \raggedright
\begin{knitrout}
\definecolor{shadecolor}{rgb}{0.969, 0.969, 0.969}\color{fgcolor}\begin{kframe}
\begin{alltt}
\hlkwd{set.seed}\hlstd{(}\hlnum{2}\hlstd{)}
\hlstd{samp} \hlkwb{<-} \hlkwd{rnorm}\hlstd{(}\hlnum{50}\hlstd{,} \hlopt{-}\hlnum{2}\hlstd{,} \hlnum{1}\hlstd{)}
\hlkwd{vs.test}\hlstd{(samp,} \hlkwc{densfun} \hlstd{=} \hlstr{'dnorm'}\hlstd{,} \hlkwc{param} \hlstd{=} \hlopt{-}\hlnum{2}\hlstd{)}
\end{alltt}

{\ttfamily\noindent\bfseries\color{errorcolor}{Error in vs.test(samp, densfun = "{}dnorm"{}, param = -2): "{}param"{}: invalid parameter (not consistent with the specified distribution)}}\end{kframe}
\end{knitrout}
}

One can estimate the p-value of the sample by Monte-Carlo simulation, even when sample size is larger than 80, by setting the optional argument \code{simulate.p.value} to \code{TRUE} (\code{NULL} by default). The number of Monte-Carlo replicates can be fixed through the optional argument~\code{B} (default is \code{B = 5000}).

\begin{knitrout}
\definecolor{shadecolor}{rgb}{0.969, 0.969, 0.969}\color{fgcolor}\begin{kframe}
\begin{alltt}
\hlkwd{set.seed}\hlstd{(}\hlnum{1}\hlstd{)}
\hlstd{samp} \hlkwb{<-} \hlkwd{rweibull}\hlstd{(}\hlnum{200}\hlstd{,} \hlkwc{shape} \hlstd{=} \hlnum{1.05}\hlstd{,} \hlkwc{scale} \hlstd{=} \hlnum{1}\hlstd{)}
\hlkwd{set.seed}\hlstd{(}\hlnum{2}\hlstd{)}
\hlkwd{vs.test}\hlstd{(samp,} \hlkwc{densfun} \hlstd{=} \hlstr{'dexp'}\hlstd{,} \hlkwc{simulate.p.value} \hlstd{=} \hlnum{TRUE}\hlstd{,} \hlkwc{B} \hlstd{=} \hlnum{10000}\hlstd{)}
\end{alltt}
\begin{verbatim}

	Vasicek-Song GOF test for the exponential distribution

data:  samp
Test statistic = 0.10907, Optimal window = 3, p-value = 0.3504
sample estimates:
   Rate 
1.15047 
\end{verbatim}
\end{kframe}
\end{knitrout}

Vasicek's estimates $V_{mn}$ are computed for all $m$ from $1$ to $n^{1/3-\delta}$, where $\delta < 1/3$; the test statistic is $I_{\widehat{m}n}$ for $\widehat{m}$ the optimal window size, as defined in~(\ref{arteqm}).
The choice of $\delta$ depends on the family of distributions of the null hypothesis. Precisely, for Weibull, Pareto, Fisher, Laplace and Beta, $\delta$ is set by default to $2/15$, while for uniform, normal, log-normal, exponential and gamma, it is set to $1/12$. These default settings result from numerous experimentations. Still, the user can choose another value through the optional argument \code{delta}.

\begin{knitrout}
\definecolor{shadecolor}{rgb}{0.969, 0.969, 0.969}\color{fgcolor}\begin{kframe}
\begin{alltt}
\hlkwd{set.seed}\hlstd{(}\hlnum{63}\hlstd{)}
\hlkwd{vs.test}\hlstd{(samp,} \hlkwc{densfun} \hlstd{=} \hlstr{'dexp'}\hlstd{,} \hlkwc{delta} \hlstd{=} \hlnum{5}\hlopt{/}\hlnum{30}\hlstd{)}
\end{alltt}
\begin{verbatim}

	Vasicek-Song GOF test for the exponential distribution

data:  samp
Test statistic = 0.16517, Optimal window = 2, p-value = 0.1538
sample estimates:
   Rate 
1.15047 
\end{verbatim}
\end{kframe}
\end{knitrout}

Note that upper-bounding the window size by $n^{1/3-\delta}$ is only required when the asymptotic normality of $I_{mn}$ is used to compute asymptotic p-values from~(\ref{asymptoticn}). When the p-value are computed by means of Monte-Carlo simulation, this upper-bound can be extended to $n/2$ by adding \code{extend = TRUE}, which may lead to a more reliable test, as illustrated below.

\begin{knitrout}
\definecolor{shadecolor}{rgb}{0.969, 0.969, 0.969}\color{fgcolor}\begin{kframe}
\begin{alltt}
\hlkwd{set.seed}\hlstd{(}\hlnum{8}\hlstd{)}
\hlstd{samp} \hlkwb{<-} \hlkwd{rexp}\hlstd{(}\hlnum{30}\hlstd{,} \hlkwc{rate} \hlstd{=} \hlnum{3}\hlstd{)}
\hlkwd{vs.test}\hlstd{(}\hlkwc{x} \hlstd{= samp,} \hlkwc{densfun} \hlstd{=} \hlstr{"dlnorm"}\hlstd{)}
\end{alltt}
\begin{verbatim}

	Vasicek-Song GOF test for the log-normal distribution

data:  samp
Test statistic = 0.30717, Optimal window = 2, p-value = 0.1206
sample estimates:
 Location     Scale 
-2.162290  1.683868 
\end{verbatim}
\end{kframe}
\end{knitrout}

\begin{knitrout}
\definecolor{shadecolor}{rgb}{0.969, 0.969, 0.969}\color{fgcolor}\begin{kframe}
\begin{alltt}
\hlkwd{vs.test}\hlstd{(}\hlkwc{x} \hlstd{= samp,} \hlkwc{densfun} \hlstd{=} \hlstr{"dlnorm"}\hlstd{,} \hlkwc{extend} \hlstd{=} \hlnum{TRUE}\hlstd{)}
\end{alltt}
\begin{verbatim}

	Vasicek-Song GOF test for the log-normal distribution

data:  samp
Test statistic = 0.3029, Optimal window = 3, p-value = 0.007
sample estimates:
 Location     Scale 
-2.162290  1.683868 
\end{verbatim}
\end{kframe}
\end{knitrout}

Enlarging the range of $m$  is also pertinent if ties are present in the sample. Indeed, the presence of ties is particularly inappropriate for performing VS tests, because some spacings $X_{(i+m)} - X_{(i-m)}$ can be null. The window size $m$ has thus to be greater than the maximal number of ties in the sample. Hence, if the upper-bound $n^{1/3-\delta}$ is less than the maximal number of ties, the test statistic can not be computed. Setting \code{extend} to \code{TRUE} can avoid this behavior, as illustrated below.

{
  \raggedright
\begin{knitrout}
\definecolor{shadecolor}{rgb}{0.969, 0.969, 0.969}\color{fgcolor}\begin{kframe}
\begin{alltt}
\hlstd{samp} \hlkwb{<-} \hlkwd{c}\hlstd{(samp,} \hlkwd{rep}\hlstd{(}\hlnum{4}\hlstd{,}\hlnum{3}\hlstd{))} \hlcom{#add ties in the previous sample}
\hlkwd{vs.test}\hlstd{(}\hlkwc{x} \hlstd{= samp,} \hlkwc{densfun} \hlstd{=} \hlstr{"dexp"}\hlstd{)}
\end{alltt}

{\ttfamily\noindent\color{warningcolor}{Warning in vs.estimate(x, densfun, ESTIM, extend, delta, relax): Ties should not be present for Vasicek-Song test}}

{\ttfamily\noindent\bfseries\color{errorcolor}{Error in vs.estimate(x, densfun, ESTIM, extend, delta, relax): Too many ties to compute Vasicek estimate.}}\end{kframe}
\end{knitrout}
}

{
  \raggedright
\begin{knitrout}
\definecolor{shadecolor}{rgb}{0.969, 0.969, 0.969}\color{fgcolor}\begin{kframe}
\begin{alltt}
\hlkwd{vs.test}\hlstd{(}\hlkwc{x} \hlstd{= samp,} \hlkwc{densfun} \hlstd{=} \hlstr{"dexp"}\hlstd{,} \hlkwc{extend} \hlstd{=} \hlnum{TRUE}\hlstd{)}
\end{alltt}

{\ttfamily\noindent\color{warningcolor}{Warning in vs.estimate(x, densfun, ESTIM, extend, delta, relax): Ties should not be present for Vasicek-Song test}}\begin{verbatim}

	Vasicek-Song GOF test for the exponential distribution

data:  samp
Test statistic = 0.025702, Optimal window = 16, p-value = 0.9052
sample estimates:
    Rate 
1.683785 
\end{verbatim}
\end{kframe}
\end{knitrout}
}

Finally, Vasicek's estimate $V_{mn}$ may exceed the parametric estimate of the entropy of the null distribution for all $m$ between $1$ and $n^{1/3-\delta}$. Then, no window size exists satisfying~(\ref{arteqm}), as illustrated below.

{
  \raggedright
\begin{knitrout}
\definecolor{shadecolor}{rgb}{0.969, 0.969, 0.969}\color{fgcolor}\begin{kframe}
\begin{alltt}
\hlkwd{set.seed}\hlstd{(}\hlnum{84}\hlstd{)}
\hlstd{ech} \hlkwb{<-} \hlkwd{rpareto}\hlstd{(}\hlnum{20}\hlstd{,} \hlkwc{mu} \hlstd{=} \hlnum{1}\hlopt{/}\hlnum{2}\hlstd{,} \hlkwc{c} \hlstd{=} \hlnum{1}\hlstd{)}
\hlkwd{vs.test}\hlstd{(}\hlkwc{x} \hlstd{= ech,} \hlkwc{densfun} \hlstd{=} \hlstr{'dpareto'}\hlstd{,} \hlkwc{param} \hlstd{=} \hlkwd{c}\hlstd{(}\hlnum{1}\hlopt{/}\hlnum{2}\hlstd{,} \hlnum{1}\hlstd{))}
\end{alltt}

{\ttfamily\noindent\bfseries\color{errorcolor}{Error in vs.estimate(x, densfun, ESTIM, extend, delta, relax): The sample entropy is greater than empirical maximal entropy for all possible window sizes; the sample may be too small or is unlikely to be drawn from the null distribution.}}\end{kframe}
\end{knitrout}
}

Enlarging the possible window sizes by setting \code{extend} to \code{TRUE} may enable Vasicek estimates to be smaller than empirical entropy.

Note that when computing the p-value by Monte-Carlo simulation, the constraint~(\ref{PKGEqnConstraint}) may not be satisfied for some replicates, whatever be the window size. These replicates are then ignored and the p-value is computed from the remaining replicates. A warning message is added to the output, informing on the number of ignored replicates.

{
  \raggedright
\begin{knitrout}
\definecolor{shadecolor}{rgb}{0.969, 0.969, 0.969}\color{fgcolor}\begin{kframe}
\begin{alltt}
\hlkwd{data}\hlstd{(contaminants)} \hlcom{#load data from package vsgoftest; see ?contaminants}
\hlkwd{set.seed}\hlstd{(}\hlnum{1}\hlstd{)}
\hlkwd{vs.test}\hlstd{(}\hlkwc{x} \hlstd{= aluminium2,} \hlkwc{densfun} \hlstd{=} \hlstr{'dpareto'}\hlstd{)}
\end{alltt}

{\ttfamily\noindent\color{warningcolor}{Warning in vs.test(x = aluminium2, densfun = "{}dpareto"{}): For 176 simulations (over 5000 ), entropy estimate is greater than empirical maximum entropy for all window sizes.}}\begin{verbatim}

	Vasicek-Song GOF test for the Pareto distribution

data:  aluminium2
Test statistic = 1.3676, Optimal window = 2, p-value < 2.2e-16
sample estimates:
         mu           c 
  0.3288148 360.0000000 
\end{verbatim}
\end{kframe}
\end{knitrout}
}

A large proportion of such ignored replicates may indicate that the original sample is too small or the null distribution does not fit it.

The function \code{vs.test} also allows to avoid the constraint~(\ref{PKGEqnConstraint}) when computing the optimal window size, by setting the optional argument \code{relax} to \code{TRUE}. This however should be used with special care, even when the p-value is computed by Monte-carlo simulation, because it may lead to spurious conclusions. Some examples will be discussed in Section~\ref{MEGOFTSecApplication}.
This option is to recover the non-parametric likelihood ratio GOF test developed by~\cite{VexlerGurevich2010_CSDA} and performed by \pkg{dbEmpLikeGOF}; see Section~\ref{PKGSecCompPkg}.

\subsection{Technical information on the internal structure of the vsgoftest package}
\label{MEGOFPSecStrPack}

\begin{figure}[ht!]
\centering
\includegraphics{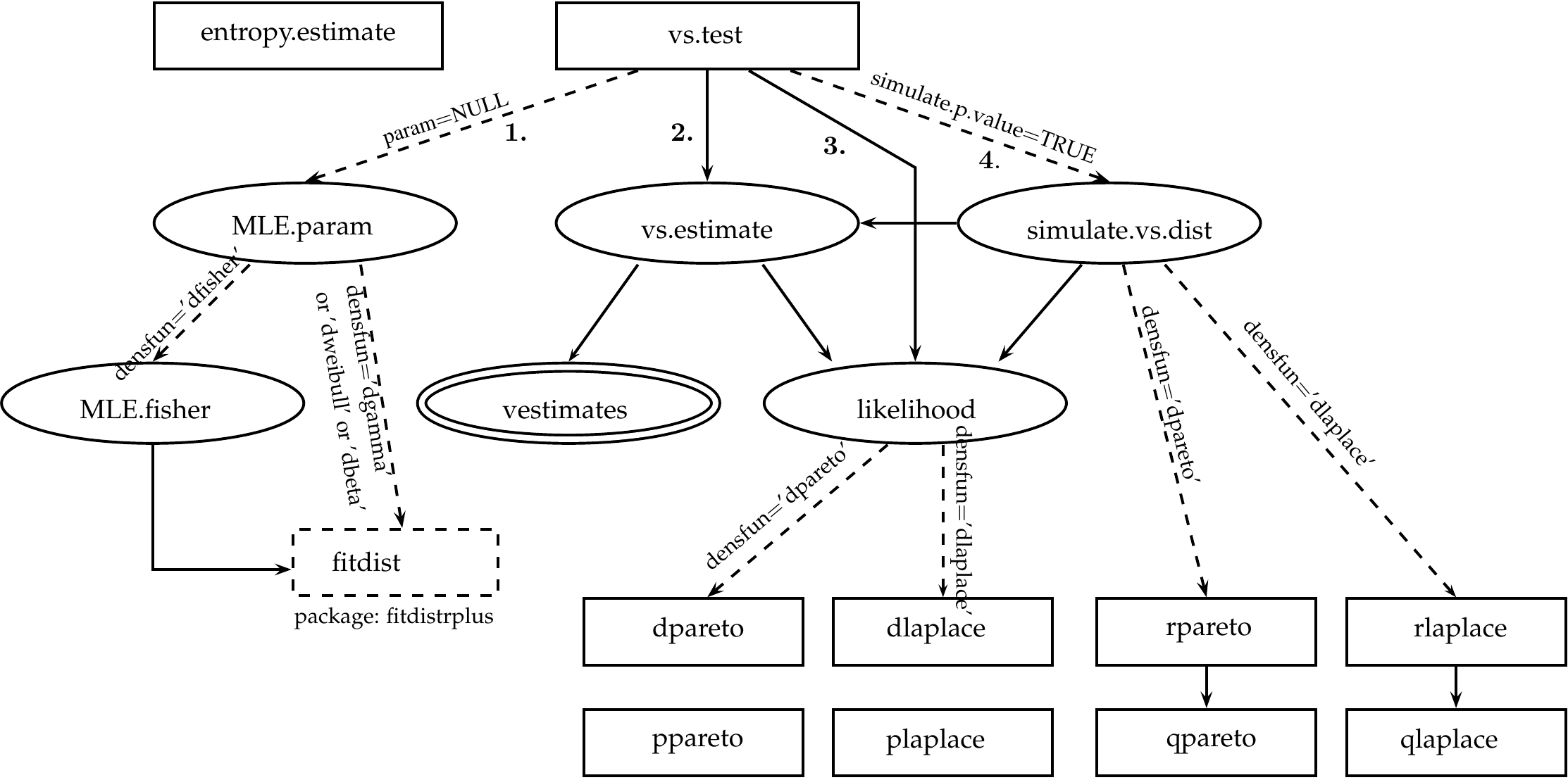}
\caption{Organisational chart of the structure of the package and connections between functions available for users (depicted by rectangles) and internal functions (depicted by ellipses). } 
\label{PKGFigOrgPkg}
\end{figure}

While \code{entropy.estimate} is a stand-alone function -- depending only on the \pkg{base} and \pkg{stats} packages, \code{vs.test} is supported by a set of internal functions -- not available for users; the structure of the package and connections between functions are described in the organisational chart presented in Figure~\ref{PKGFigOrgPkg}. Functions available for users are depicted by rectangles while internal functions are depicted by ellipses.
An arrow connecting a function to another means that the first function (say master function) calls the second (slave) during execution. When such a call is optional (depending on arguments given in the master function), the arrow is dashed and annotated with the corresponding argument settings. The function \code{fitdist} depicted by a dashed rectangle is a function implemented in the \pkg{fitdistrplus} package~\cite{delignette}. The double-lined ellipse depicts a \proglang{C++} encoded function that has been integrated via the package \pkg{Rcpp}~\cite{RcppEddelbuettel}.

The \pkg{vsgoftest} package is structured in such a way so as to:
\begin{itemize}
 \item Allow easy access to the code source. Especially, the master function \code{vs.test} calls four slave functions corresponding to the following tasks (enumerated according to the organisational chart of Figure~\ref{PKGFigOrgPkg}):
 \begin{enumerate}
  \item computing the MLE of the parameter $\theta$ of the null distribution through the function \code{MLE.param}.
  \item Computing Vasicek estimate $V_{\widehat{m}n}$ of Shannon entropy for the sample with the optimal window $\widehat{m}$ given by~(\ref{arteqm}).
  \item Computing the VS test statistic $I_{\widehat{m}n}$.
  \item Computing the p-value associated to the sample. If the sample size is either greater than $80$ or the optional argument \code{simulate.p.value} is \code{TRUE}, then the p-value is estimated by means of Monte-Carlo simulation performed by the internal function \code{simulate.vs.dist}.
 \end{enumerate}
 \item Limit dependence to other packages. In this aim, density, cumulative density and quantile functions as well as random generators for Pareto and Laplace distributions have been encoded, even if they are available in other \proglang{R} packages such as \pkg{VGAM} in~\cite{VGAM_package}, \pkg{POT} in~\cite{POT_package} and~\pkg{smoothmest} in~\cite{smoothmest}.
The MLE $\widehat{\theta}_n$ of the parameter of the null distribution is computed thanks to the function \code{fitdist} of the \pkg{fitdistrplus} package only if no closed form expression is known for it, \textit{i.e.}, for Gamma, Weibull, Beta and Fisher distributions. Otherwise, the closed form expression is used.
 \item Optimize time and resources, especially for Monte-Carlo simulation. To this end, the most time-consuming part of the procedure -- namely, the computation of Vasicek estimate for all possible window sizes, has been converted to \proglang{C++} and integrated to the package via \pkg{Rcpp}, in the internal function \code{vestimates}.
\end{itemize}

\section{Performance of Vasicek-Song tests} \label{MEGOFTSecComparison}

First, a review of power studies of VS tests available in literature is presented in Section~\ref{seccomputation}. Then, power comparisons of VS tests and classical GOF tests are proposed when applied to discriminate between close distributions, such as Pareto versus shifted log-normal and Exponential versus Weibull. 
Finally, the features of packages \pkg{vsgoftest} and \pkg{dbEmpLikeGOF} are compared in Section~\ref{PKGSecCompPkg}; the methodological differences are highlighted, the higher performance of \pkg{vsgoftest} both in terms of power and computational time is pointed out and illustrated.

 \subsection{Power computation}
 \label{seccomputation}

Comparisons of the power properties of VS tests are widely discussed in the literature. Various choices of null and alternative distribution families are considered. VS tests are shown to generally outperform classical GOF tests. A comprehensive list of these references is given in this section, with main conclusions summarized in Table~\ref{tab:powercomparisons}. 
Especially, 
power properties of the VS test for normality have been discussed by \citet{vasicek}, 
\citet{arizono} and \citet{gurevich} among many others. Compared with many tests, including Kolmogorov-Smirnov (KS), Cram{\'e}r-von Mises (CvM), Anderson-Darling (AD) and Shapiro-Wilk (SW), the VS test exhibits higher power for most of alternative distributions. 
When the null distribution is an exponential distribution, the VS test is also shown in \citet{ebrahimi} to be more powerful than the Van-Soest and Finkelstein and Schafer tests, which are modified versions of respectively CvM and KS tests, for various alternative distributions such as Weibull, gamma and log-normal. 
\cite{choikim} for Laplace and \cite{lequesne_these} for Pareto  show that the VS test is more powerful than EDF tests, for various alternative distributions. The uniform VS GOF test is shown to outperform many other tests for alternative distributions having most of their mass near 0.5, but remains less powerful than CvM and Watson tests for other alternative distributions.

\begin{table}[!ht]
  \centering
 \scalebox{0.9}{
  \footnotesize
  \begin{tabular}{|p{3cm}|l|p{4cm}|p{3cm}|p{3cm}|}
  \hline 
  \textbf{Reference} & \textbf{Null distrib.}  &  \textbf{Alt. distrib.} & \textbf{VS compared with} & \textbf{Most powerful} \\ \hline \hline 
  \cite{vasicek} & \multirow{3}{*}{Normal} & exponential, gamma, uniform, beta, Cauchy & KS, CvM, Kuiper, Watson, AD, SW & AD (for Cauchy), VS (for others) \\ \cline{1-1} \cline{3-5} 
  \cite{arizono} & & log-normal, uniform, $\chi^2$, student & KS, CvM, $\chi^2 $ & VS  \\ \cline{1-1} \cline{3-5} 
  \cite{gurevich} & & log-normal, $\chi^2$, Student, uniform, exponential, gamma, beta, Cauchy  & KS & KS (for Student and Cauchy), VS (for others)\\ \hline 
  \cite{ebrahimi} & Exponential & Weibull, gamma, log-normal & Van-Soest, Finkelstein and Schafer & VS \\ \hline 
  \cite{dudewicz} & Uniform & Distributions defined on $[0,1]$ & KS, CvM, Kuiper, Watson, AD, log-statistic, $\chi^2$ & VS (for alternative having most of its mass near 0.5), CvM or Watson (for others)\\ \hline 
  \cite{choikim} & Laplace & Normal, Student, logistic, Cauchy, uniform, chi-squared, Weibull, log-normal, extreme value and inverse Gaussian & KS, CvM, AD, Kuiper and Watson & VS  \\ \hline
  \cite{lequesne_these} & Pareto & Weibull, gamma, log-normal, two-parameter exponential & KS, AD & VS \\ \hline \hline
  \cite{mudholkar} & Inv. Gaussian & exponential, uniform, Weibull and log-normal &  KS & VS (uniform and Weibull)\\ \hline
  \cite{alizadeh} & Rayleigh & Weibull, gamma, log-normal, half-normal, uniform, modified extreme value, linear increasing failure rate law, Dhillon's law and Chen's distribution  & KS, CvM, AD, Kuiper and Watson & VS (for uniform) and AD (for other alternatives) \\ \hline
  \cite{perez} & Gumbel & Weibull, log-normal, normal, logistic, Cauchy, Student, gamma and Fr\'echet  & KS, CvM, AD, Kuiper and Kinnison & AD (for heavy tails), VS (for others) \\ \hline
  \cite{tsujitani} & Extreme-Value & Normal and 3-parameter log-Weibull & KS, CvM, AD, Kuiper, Mann et modified Mann test & VS\\ \hline
  \cite{lund} & von Mises & Mixtures of Von Mises distributions (bimodal, skewed, long-tailed and half), the cardoid and triangular distributions  & Watson and integrated squared error test & Watson (for long-tailed), VS (for half) and fairly equal for other alternatives\\ \hline
  \end{tabular}
  }
  \caption{Power studies of VS tests performed in the literature.}
  \label{tab:powercomparisons}
\end{table}

On the basis of power computation in the literature, we choose to compare the power of the VS test to the KS, CvM and AD tests, for close null and alternative distributions. In particular, difficulties in distinguishing a Pareto tail from that of a log-normal is an issue; see for example \citet{malevergne}. For illustration, we estimate through Monte-Carlo simulation the power of VS, KS, CvM and AD of Pareto distributions applied to samples drawn from a (shifted) log-normal distribution. We simulate $10000$ replicates $x_{1}^n$ of a random sample $X_1^n$ drawn from a shifted log-normal distribution $\mathcal{LN}(0,\sigma)$ with support~$[1,\infty[$ and $\sigma=1, 1.25$, for $n \in \{20, 30, 50, 100\}$. 
Then, we apply the tests for the simple null hypothesis $H_0: P=\mathcal{P}ar(1,\mu)$ , for $\mu=1$ when $\sigma =1$ and $\mu=0.8$ when $\sigma =1.25$; the power is estimated by the proportion of rejections of the null hypothesis among the $10000$ replicates. 
The following code chunk illustrates the procedure, for $\sigma = \mu =1$ and $n = 20$, using the VS test. 
This procedure immediately adapts to other values of $\sigma$, $\mu$ and $n$ and to other tests\footnote{The seed of the pseudo-random number generator has been changed for each couple of $\mu$ and $n$; the whole procedure yielding Table~\ref{tab:power} is available in the file \textit{vsgoftest\_performances.R}, in the directory \textit{inst/doc} of the package source file.}. Results are presented in Table~\ref{tab:power} (top). 

\begin{knitrout}
\definecolor{shadecolor}{rgb}{0.969, 0.969, 0.969}\color{fgcolor}\begin{kframe}
\begin{alltt}
\hlstd{N} \hlkwb{<-} \hlnum{10000}
\hlstd{n} \hlkwb{<-} \hlnum{20}
\hlstd{mu} \hlkwb{<-} \hlnum{1}
\hlkwd{set.seed}\hlstd{(}\hlnum{54}\hlstd{)}
\hlstd{res.pow} \hlkwb{<-} \hlkwd{replicate}\hlstd{(}\hlkwc{n} \hlstd{= N,}
                     \hlkwc{expr} \hlstd{=} \hlkwd{vs.test}\hlstd{(}\hlkwc{x} \hlstd{=} \hlnum{1} \hlopt{+} \hlkwd{rlnorm}\hlstd{(n,}
                                                   \hlkwc{meanlog} \hlstd{=} \hlnum{0}\hlstd{,}
                                                   \hlkwc{sdlog} \hlstd{=} \hlnum{1}\hlstd{),}
                                    \hlkwc{densfun} \hlstd{=} \hlstr{'dpareto'}\hlstd{,}
                                    \hlkwc{param} \hlstd{=} \hlkwd{c}\hlstd{(}\hlnum{1}\hlstd{,}\hlnum{1}\hlstd{),}
                                    \hlkwc{simulate.p.value} \hlstd{=} \hlnum{TRUE}\hlstd{,}
                                    \hlkwc{B} \hlstd{=} \hlnum{1000}\hlstd{)}\hlopt{$}\hlstd{p.value)}
\end{alltt}
\end{kframe}
\end{knitrout}

The power of VS, KS, CvM and AD tests is similarly computed for null exponential and alternative Weibull distributions. The Weibull distribution $\mathcal{W}(a,b)$ reduces to an exponential distribution $\mathcal{E}(1/a)$ when $b=1$. The main aim is thus to determine which test better discriminates between these distributions when the shape parameter of the Weibull distribution is close to $1$, precisely $b=1.2$ and $b=1.3$. Results are given in Table~\ref{tab:power} (bottom), clearly showing that the VS test outperforms EDF tests.

 \begin{table}[t]
   \small
  \centering
   \begin{tabular}{|c|c|c|c|c||c|c|c|c|}
   \hline
   &  VS & KS & AD & CvM &  VS & KS & AD & CvM\\
    \hline
   \hline
  & \multicolumn{4}{c||}{\tiny $H_0 : \mathcal{P} (1,1) ;  H_1: 1+\mathcal{LN}(0,1)$} & \multicolumn{4}{c|}{\tiny $H_0:  \mathcal{P} (1,0.8) ; H_1: 1+ \mathcal{LN}(0,1.25)$}\\
   \hline
   n=20 & 59.79 & 8.93 & 6.72 & 7.02 & 40.62 & 16.36 & 13.44 & 16.21\\
   n=30 & 77.66 & 15.79 & 22.83 & 16.61 & 55.50 & 27.05 & 26.54 & 26.98\\
   n=50 & 94.02 & 37.39 & 68.02 & 46.86 & 76.83 & 50.70 & 58.87 & 52.36\\
   n=100 & 99.99 & 85.90 & 99.83 & 96.36 & 98.42 & 89.22 & 97.58 & 92.07\\
   \hline \hline
& \multicolumn{4}{c||}{\tiny $H_0:\mathcal{E}(1/2) ; H_1:\mathcal{W}(1.2,2)$} &  \multicolumn{4}{c|}{\tiny $H_0:\mathcal{E}(1/2) ; H_1:\mathcal{W}(1.3,2)$}\\
    \hline
   n=20 & 9.97 & 5.06 & 3.65 & 4.63 & 14.67 & 5.27 & 3.28 & 4.41\\
   n=30 & 12.05 & 6.10 & 4.29 & 5.19 & 19.93 & 7.40 & 5.65 & 6.45\\
   n=50 & 13.47 & 7.37 & 6.50 & 6.80 & 25.86 & 11.28 & 11.30 & 10.53\\
   n=100 & 25.23 & 11.35 & 14.04 & 11.42 & 67.14 & 21.91 & 34.67 & 24.60\\
   \hline
   \end{tabular}
   \caption{Power (expressed as percentage of true rejection) of VS, KS, CvM and AD tests for testing Pareto and exponential distributions against shifted log-normal (top) and Weibull distributions (bottom).}
\label{tab:power}

 \end{table}

Note that the above procedure for comparing the power of GOF tests adapts easily to other sets of null and alternative distributions.

\subsection{vsgoftest versus dbEmpLikeGOF for testing uniformity and normality} \label{PKGSecCompPkg}

As mentioned in the introduction section, 
The package \pkg{dbEmpLikeGOF} in~\cite{dbEmpLikeGOF_package} performs uniformity and normality tests based on empirical likelihood ratios (ELR)  -- say ELR tests. These tests are strongly linked to VS tests. 
Precisely, for testing the normality of a sample $X_1, \dots X_n$, the ELR test statistic is $\log V_n$, where
$$V_n = \min_{1 \leq m < n^{1/2}} \left[ (2\pi e \check{S}^2)^{n/2} \prod_{i=1}^n \frac{2m}{n \left[ X_{(i+m) - X_{(i-m)}} \right]} \right] \ \textrm{and} \ \check{S}^2 = \frac{1}{n-1} \sum_{i=1}^n \left(X_i- \frac{1}{n}\sum_{j=1}^n X_j \right)^2.$$
Mere algebra yields
\begin{equation} \label{PKGEqnLinkTeststats}
\log V_n = n I_{\widetilde{m}n} + \frac{1}{2},
\end{equation}
with $\widetilde{m} \in \arg \!\! \max_{1 \leq m < n^{1/2}} V_{mn}$.
Hence, ELR and VS tests differ only in the window size choice: the upper bound is  $n^{1/2}$ for the ELR test while it is (by default) $n^{1/4}$ for the VS test and the constraint~(\ref{PKGEqnConstraint}) is not taken into account by the EL test. 
Enlarging the upper bound from $n^{1/4}$ to $n^{1/2}$ may lead to a more powerful decision rule, as mentioned and illustrated in Section~\ref{MEGOFPSecVSTest}. 
Still, practically, the normality VS test tends to outperform the ELR test when applied to heavy tailed samples, as illustrated by Table~\ref{PKGTableCompVSELR}\footnote{The comparison procedure is available in the file \textit{vsgoftest\_performances.R}, in the directory \textit{inst/doc} of the package source file.}.
Moreover, the upper bound $n^{1/4}$ legitimates the use of the asymptotic distribution of $I_{\widehat{m}n}$ in \code{vs.test}, which is not performed by \code{dbEmpLikeGOF}. 
As previously mentioned in Section~\ref{MEGOFPSecVSTest}, disabling the constraint~(\ref{PKGEqnConstraint}) may lead to spurious conclusions. 

\begin{table}
 \centering
 \begin{tabular}{|r||c|c||c|c|}
  \hline
  & VS & ELR & VS & ELR \\ \hline 
  & \multicolumn{2}{|c||}{$H_1 : \mathcal{L}(0,1)$} & \multicolumn{2}{|c|}{$H_1 : \mathcal{S}(4)$} \\ \hline
  $n=50$ & 17.8 & 16.0 &  16.1 & 14.6 \\ \hline
  $n=200$ & 86.2 & 64.9 & 71.2 & 35.8 \\ \hline
  \end{tabular}
  \caption{Power comparisons between VS and ELR normality tests for samples drawn from  a Laplace distribution (left) and a Student distribution (right). Power is estimated by means of Monte-Carlo simulation based on 1000 replicates of the samples.}
  \label{PKGTableCompVSELR}
\end{table}

The ELR test can be performed using \code{vs.test}, by suitably setting its arguments, as illustrated by the following code chunk
\footnote{Some slight difference remains between the two computed values, due to numerical inaccuracy in computation procedures: the estimated entropy of the null distribution is computed from the closed form expression~(\ref{MEGOFPEqnEntNormal}) in \code{dbEmpLikeGOF} while it is computed as the empirical mean of the log-likelihood of the sample in \code{vs.test}.}.

\begin{knitrout}
\definecolor{shadecolor}{rgb}{0.969, 0.969, 0.969}\color{fgcolor}\begin{kframe}
\begin{alltt}
\hlkwd{set.seed}\hlstd{(}\hlnum{1}\hlstd{)}
\hlstd{samp} \hlkwb{<-} \hlkwd{rnorm}\hlstd{(}\hlnum{50}\hlstd{)}
\hlstd{res.vs} \hlkwb{<-} \hlkwd{vs.test}\hlstd{(}\hlkwc{x} \hlstd{= samp,} \hlkwc{densfun} \hlstd{=} \hlstr{'dnorm'}\hlstd{,} \hlkwc{delta} \hlstd{=} \hlopt{-}\hlnum{1}\hlopt{/}\hlnum{6}\hlstd{,} \hlkwc{relax} \hlstd{=} \hlnum{TRUE}\hlstd{)}
\hlstd{res.vs}\hlopt{$}\hlstd{statistic}\hlopt{*}\hlnum{50} \hlopt{+}\hlnum{1}\hlopt{/}\hlnum{2}
\end{alltt}
\begin{verbatim}
Test statistic 
      7.970748 
\end{verbatim}
\begin{alltt}
\hlkwd{library}\hlstd{(dbEmpLikeGOF)}
\hlstd{res.el} \hlkwb{<-} \hlkwd{dbEmpLikeGOF}\hlstd{(}\hlkwc{x} \hlstd{= samp,} \hlkwc{testcall} \hlstd{=} \hlstr{'normal'}\hlstd{,} \hlkwc{vrb} \hlstd{=} \hlnum{FALSE}\hlstd{)}
\hlstd{res.el}\hlopt{$}\hlstd{teststat}
\end{alltt}
\begin{verbatim}
[1] 7.975815
\end{verbatim}
\end{kframe}
\end{knitrout}

\begin{figure}[ht!]
 \centering
 \begin{tabular}{c}
  \includegraphics[angle = 270, width = 0.6\textwidth]{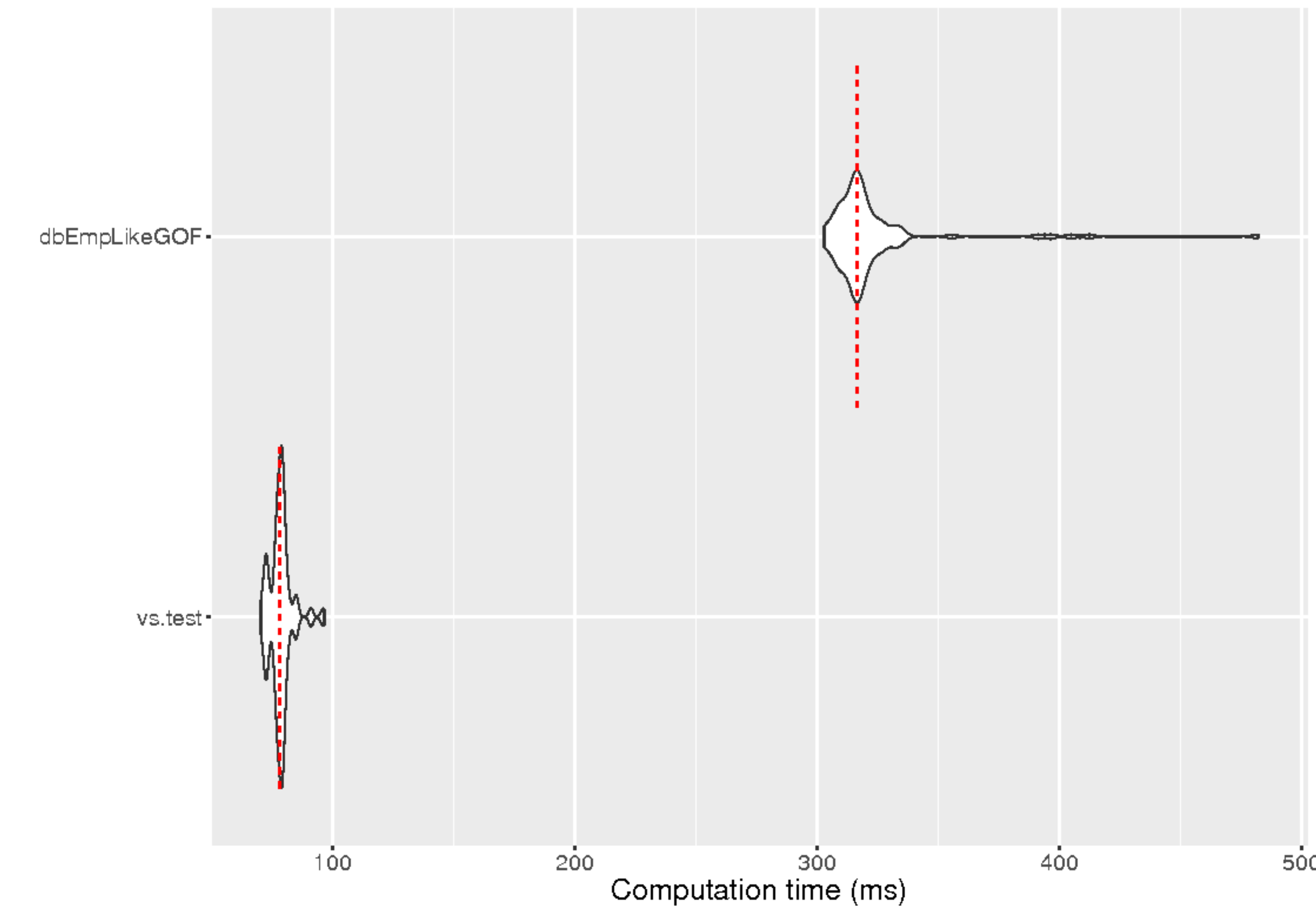} \\
  \includegraphics[angle = 270, width = 0.6\textwidth]{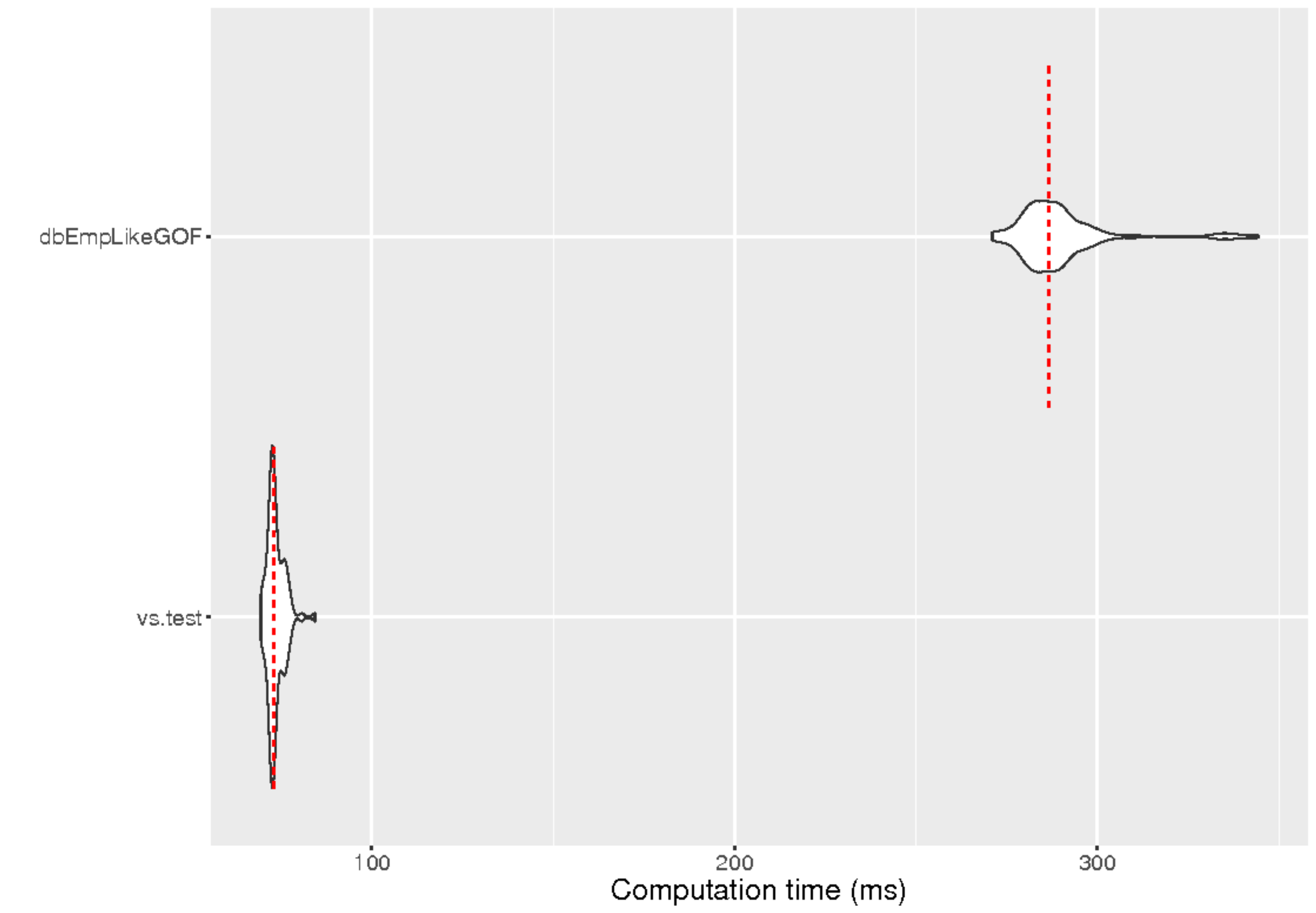}
 \end{tabular}
 \caption{Distributions (violin plots) of computation time of 100 iterations of the functions dbEmpLikeGOF (from package \pkg{dbEmpLikeGOF}) and vs.test for testing normality of a sample of size 50 drawn from the standard normal distribution (top) and for testing uniformity of a sample of size 50 drawn from the uniform distribution on $[0,1]$ (bottom). The p-values are estimated by Monte-Carlo simulation, based on 1000 replicates. The red dashed lines are the median computation times.}
 \label{PKGFigCompTime}
\end{figure}

Additionnaly, from a computational view point, some significant differences exist between the functions \code{dbEmpLikeGOF} and \code{vs.test}. Precisely, the p-value returned by \code{dbEmpLikeGOF} is computed by default by linear interpolation from a table of pre-computed p-values for various sample sizes (from $n=10$ to $10000$) and test statistic values; the p-value can be approximated by Monte-Carlo simulation by setting \code{pvl.Table = FALSE}. By default, \code{vs.test} computes the p-value by means of Monte-Carlo simulation or uses the asymptotic distribution~(\ref{asymptoticn2}), depending on the sample size. In both cases, \code{vs.test} is approximately five times faster than \code{dbEmpLikeGOF}, as illustrated by Figure~\ref{PKGFigCompTime}\footnote{Simulations have been performed on a Dell Lattitude E5580 laptop, equipped with an Intel$^{\mbox{{\tiny\textregistered}}}$ Core{\scriptsize\texttrademark} i7-7600U CPU at 2.80GHz x 4, with 16GB RAM.  \proglang{R} code for generating these figures is available in the file
\textit{vsgoftest\_performances.R}, in the directory \textit{doc} of the package source file.}.

\section{Application to real data} \label{MEGOFTSecApplication}

The \pkg{vs.test} package contains environmental data originating from a guidance report edited by the Technology Support Center of the United States Environmental Protection Agency; see~\cite{singh}. According to~\cite{singh}, environmental scientists take remediation decisions at suspected sites based on organic and inorganic contaminant concentration measurements. These decisions usually derive from the computation of confidence upper bounds for contaminant concentrations. Testing the goodness-of-fit of specified models hence appears of prior interest. \cite{singh} also points out that contaminant concentration data from sites often appear to follow a skewed probability distribution, making the log-normal family a frequently-used model. The authors illustrate their purpose by applying Shapiro-Wilk test to the log-transformed of the samples \code{aluminium1}, \code{manganese}, \code{aluminium2} and \code{toluene} (stored in the present package)\footnote{A succinct description of these data is available by executing the following \proglang{R} command: \code{?contaminants}}; see the empirical skewness computed in the following chunk.

\begin{knitrout}
\definecolor{shadecolor}{rgb}{0.969, 0.969, 0.969}\color{fgcolor}\begin{kframe}
\begin{alltt}
\hlkwd{data}\hlstd{(contaminants)} \hlcom{#Load environmental data from package}
\hlcom{#Package DescTools required for this chunk}
\hlkwd{unlist}\hlstd{(}\hlkwd{lapply}\hlstd{(}\hlkwc{X} \hlstd{=} \hlkwd{list}\hlstd{(aluminium1, manganese, aluminium2, toluene),}
       \hlkwc{FUN} \hlstd{= DescTools}\hlopt{::}\hlstd{Skew))}
\end{alltt}
\begin{verbatim}
[1] 2.323343 1.698686 1.996607 3.961129
\end{verbatim}
\end{kframe}
\end{knitrout}

The following code chunks intend to illustrate the use and behavior of the function \code{vs.test} for these environmental data. The significant level is fixed to $0.1$ as in \cite{singh}. 
Note that warning messages notifying that there are ties in the samples have been dropped out from outputs.

\begin{knitrout}
\definecolor{shadecolor}{rgb}{0.969, 0.969, 0.969}\color{fgcolor}\begin{kframe}
\begin{alltt}
\hlkwd{set.seed}\hlstd{(}\hlnum{1}\hlstd{)}
\hlkwd{vs.test}\hlstd{(}\hlkwc{x} \hlstd{= aluminium1,} \hlkwc{densfun} \hlstd{=} \hlstr{'dlnorm'}\hlstd{)}
\end{alltt}
\begin{verbatim}

	Vasicek-Song GOF test for the log-normal distribution

data:  aluminium1
Test statistic = 0.31232, Optimal window = 2, p-value = 0.3372
sample estimates:
Location    Scale 
6.225681 1.609719 
\end{verbatim}
\end{kframe}
\end{knitrout}

The log-normal hypothesis is not rejected for \code{aluminium1}. Similar results are obtained for \code{manganese}. Log-normality is rejected for \code{aluminium2}.

\begin{knitrout}
\definecolor{shadecolor}{rgb}{0.969, 0.969, 0.969}\color{fgcolor}\begin{kframe}
\begin{alltt}
\hlkwd{set.seed}\hlstd{(}\hlnum{1}\hlstd{)}
\hlkwd{vs.test}\hlstd{(}\hlkwc{x} \hlstd{= aluminium2,} \hlkwc{densfun} \hlstd{=} \hlstr{'dlnorm'}\hlstd{)}
\end{alltt}
\begin{verbatim}

	Vasicek-Song GOF test for the log-normal distribution

data:  aluminium2
Test statistic = 0.48369, Optimal window = 2, p-value = 0.0256
sample estimates:
 Location     Scale 
8.9273293 0.8264409 
\end{verbatim}
\end{kframe}
\end{knitrout}

Due to numerous ties in \code{toluene}, \code{vs.test} can not compute Vasicek entropy estimate unless \code{extend} is set to \code{TRUE}. Still, \code{vs.test} notifies that the constraint~(\ref{PKGEqnConstraint}) is violated for all window sizes, which suggests that data are not likely to be drawn from the log-normal distribution; see Section~\ref{MEGOFPSecFramework}.
Turning \code{relax} to \code{TRUE} yields the following result.

\begin{knitrout}
\definecolor{shadecolor}{rgb}{0.969, 0.969, 0.969}\color{fgcolor}\begin{kframe}
\begin{alltt}
\hlkwd{set.seed}\hlstd{(}\hlnum{1}\hlstd{)}
\hlkwd{vs.test}\hlstd{(}\hlkwc{x} \hlstd{= toluene,} \hlkwc{densfun} \hlstd{=} \hlstr{'dlnorm'}\hlstd{,} \hlkwc{extend} \hlstd{=} \hlnum{TRUE}\hlstd{,} \hlkwc{relax} \hlstd{=} \hlnum{TRUE}\hlstd{)}
\end{alltt}
\begin{verbatim}

	Vasicek-Song GOF test for the log-normal distribution

data:  toluene
Test statistic = -2.4984, Optimal window = 11, p-value = 0.7308
sample estimates:
Location    Scale 
4.651002 3.579041 
\end{verbatim}
\end{kframe}
\end{knitrout}

Again, this last result looks spurious because the test statistic is negative -- resulting from (\ref{PKGEqnConstraint}) not being satisfied by setting \code{relax = TRUE}. An alternative is to test normality of the log-transformed sample as follows.

\begin{knitrout}
\definecolor{shadecolor}{rgb}{0.969, 0.969, 0.969}\color{fgcolor}\begin{kframe}
\begin{alltt}
\hlkwd{set.seed}\hlstd{(}\hlnum{1}\hlstd{)}
\hlkwd{vs.test}\hlstd{(}\hlkwc{x} \hlstd{=} \hlkwd{log}\hlstd{(toluene),} \hlkwc{densfun} \hlstd{=}\hlstr{'dnorm'}\hlstd{,} \hlkwc{extend} \hlstd{=} \hlnum{TRUE}\hlstd{)}
\end{alltt}
\begin{verbatim}

	Vasicek-Song GOF test for the normal distribution

data:  log(toluene)
Test statistic = 0.6536, Optimal window = 11, p-value = 2e-04
sample estimates:
    Mean St. dev. 
4.651002 3.579041 
\end{verbatim}
\end{kframe}
\end{knitrout}

The log-normal hypothesis is not rejected for \code{aluminium1} and \code{manganese} while it is rejected for \code{aluminium2} and \code{toluene}. These results are consistent with those obtained by \cite{singh}. Further, the goodness-of-fit to the Pareto distributions is performed for \code{aluminium2} and \code{toluene}. Log-normal and  Pareto distributions usually compete with closely related generating processes and hard to distinguish tail properties; see for example \citet{malevergne}. Goodness-of-fit of Pareto distribution is rejected for \code{aluminium2}.

\begin{knitrout}
\definecolor{shadecolor}{rgb}{0.969, 0.969, 0.969}\color{fgcolor}\begin{kframe}
\begin{alltt}
\hlkwd{set.seed}\hlstd{(}\hlnum{1}\hlstd{)}
\hlkwd{vs.test}\hlstd{(}\hlkwc{x} \hlstd{= aluminium2,} \hlkwc{densfun} \hlstd{=} \hlstr{'dpareto'}\hlstd{)}
\end{alltt}
\begin{verbatim}

	Vasicek-Song GOF test for the Pareto distribution

data:  aluminium2
Test statistic = 1.3676, Optimal window = 2, p-value < 2.2e-16
sample estimates:
         mu           c 
  0.3288148 360.0000000 
\end{verbatim}
\end{kframe}
\end{knitrout}

Applying \code{vs.test} to \code{toluene} with default settings yields no result because of numerous ties and the violation of~(\ref{PKGEqnConstraint}).
Uniformity of the sample transformed by the cumulative density function of the Pareto distribution can be tested as follows. Goodness-of-fit  of the Pareto distribution is not rejected for \code{toluene}.

\begin{knitrout}
\definecolor{shadecolor}{rgb}{0.969, 0.969, 0.969}\color{fgcolor}\begin{kframe}
\begin{alltt}
\hlcom{#Compute the MLE of parameters of Pareto dist.}
\hlstd{res.test} \hlkwb{<-} \hlkwd{vs.test}\hlstd{(}\hlkwc{x} \hlstd{= toluene,}
                    \hlkwc{densfun} \hlstd{=} \hlstr{'dpareto'}\hlstd{,}
                    \hlkwc{extend} \hlstd{=} \hlnum{TRUE}\hlstd{,} \hlkwc{relax} \hlstd{=} \hlnum{TRUE}\hlstd{)}
\hlcom{#Test uniformity of transformed data}
\hlkwd{set.seed}\hlstd{(}\hlnum{5}\hlstd{)}
\hlkwd{vs.test}\hlstd{(}\hlkwc{x} \hlstd{=} \hlkwd{ppareto}\hlstd{(toluene,}
                    \hlkwc{mu} \hlstd{= res.test}\hlopt{$}\hlstd{estimate[}\hlnum{1}\hlstd{],}
                    \hlkwc{c} \hlstd{= res.test}\hlopt{$}\hlstd{estimate[}\hlnum{2}\hlstd{]),}
        \hlkwc{densfun} \hlstd{=}\hlstr{'dunif'}\hlstd{,} \hlkwc{param} \hlstd{=} \hlkwd{c}\hlstd{(}\hlnum{0}\hlstd{,}\hlnum{1}\hlstd{),} \hlkwc{extend} \hlstd{=} \hlnum{TRUE}\hlstd{)}
\end{alltt}
\begin{verbatim}

	Vasicek-Song GOF test for the uniform distribution with Min=0,
	Max=1

data:  ppareto(toluene, mu = res.test$estimate[1], c = res.test$estimate[2])
Test statistic = 0.25383, Optimal window = 10, p-value = 0.2496
\end{verbatim}
\end{kframe}
\end{knitrout}

\section*{Conclusion}

Vasicek-Song tests constitute powerful GOF tests for classical parametric families of distributions, relying on an information theoretical  framework.  They can be easily performed by using the \pkg{vsgoftest} package for \proglang{R}. Default and optional settings of the functions provided by the package make the procedure both intuitive and flexible. Its application to real datasets manages to illustrate its practical usage. 

The package allows for testing GOF of a significant list of parametric models; this list could be extended in further releases. New entropy-based GOF tests could also be considered by using Rényi entropy and divergence --  see~\cite{lequesne_maxent2}, thus extending even more the class of possible distributions, e.g., Student distributions.

\end{document}